\documentclass[12pt]{iopart}

\pdfoutput=1

\usepackage{graphicx,sidecap}
\usepackage{bm}
\usepackage{braket}
\usepackage{amssymb}
\usepackage{mathrsfs}
\usepackage{amsmath}
\usepackage{xcolor}
\usepackage{hyperref}
\usepackage{enumerate}
\usepackage[toc,title]{appendix}
\usepackage[margin=30pt, bf, font=footnotesize, center, justification=justified]{caption}[2004/07/16]

\hypersetup{colorlinks,bookmarksopen,bookmarksnumbered,
citecolor=red,
linkcolor=blue,
pdfstartview=green,
urlcolor=purple}

\catcode`@=11 
\renewcommand\tableofcontents{%
  \section*{\contentsname}%
  \@starttoc{toc}%
}
\catcode`@=12

\def\be{\begin{equation}}
\def\ee{\end{equation}}

\def\bea{\begin{eqnarray}}
\def\eea{\end{eqnarray}}

\begin{document}

\title[Quantum Renyi relative entropies on a spin chain with a interface defects
]
{Quantum  Renyi relative entropies on a spin chain with interface defects}

\vspace{.5cm}

\begin{center}
\author{Ra\'ul Arias}
\end{center}
\address{SISSA and INFN Sezione di Trieste, via Bonomea 265, 34136 Trieste, Italy.}

\vspace{.5cm}

\begin{abstract}
We compute the quantum Renyi relative entropies in an infinite spinless fermionic chain with a defect. Doing a numerical analysis we will show that the resulting quantity depends non trivially on the effective central charge of the theory. Moreover, we will see that an explicit analytic expression can be written for all of them and from that one can read the quantum fidelity and the relative entropy.
\end{abstract}

\maketitle

\section{Introduction}

From the last years there was a big effort on the community to merge quantum information quantities with condensed matter systems and high energy physics. The most famous and very well study of these quantities is the entanglement entropy, which has a lot of applications, for example it can characterize topological quantum phase transitions or can be used to compute the central charge along a renormalization group flow in Quantum field theory (see \cite{pasqualerev, casinirev} for good reviews). But, less is known about quantum Renyi relative entropies (QRRE), a family of quantities that interpolate between the quantum fidelity and the relative entropy. The last two quantities measures distances between two states, or in other words measure how distinguishable they are. For example, the probability (p) to confound two states $\rho_1$ and $\rho_2$  after a large number $N$ of measurementes decays with the realtive entropy as $p\sim e^{-N S(\rho_1|| \rho_2)}$ \cite{relativereview}.
Some previous works studying QRRE on conformal field theories using the replica trick are \cite{nima1, alice}.

Many applications of the fidelity and relative entropy can be mentioned but we want to stress just a couple of them to exemplify its importance in different branches of physics. In one hand the quantum fidelity can be used to find the location of a quantum critical point in a system which have quantum phase transition. In fact, measuring the fidelity between two states with different values of the order parameter, $g$, one can see a peaked minimum when one state is for $g<g_c$ and the second state has $g>g_c$ (see \cite{gu, zanardi} for a review). In other hand the relative entropy can be used to give a good definition of the local temperatures related to the modular flow in quantum field theory \cite{Arias} or for example to give sense to the Bekenstein bound in black hole physics \cite{horaciobeke}. In \cite{ugajinnume} the relative entropy was computed numerically for the XXZ chain at criticality between primary states of the underlying conformal field theory and recently it was computed also between primary states of $1+1$ dimensional CFT \cite{paola} and in in-homogeneous quantum systems \cite{sara}. Although all of these computations were done using replica trick.

In a different setup, the QRRE were studied using the tomita-takesaki theorem in Algebraic Quantum Field Theory \cite{nima}. 
A more recent work, \cite{horacio}, showed that the QRRE between two states in a boundary conformal field theory that describes some aspects of the Kondo model can be bounded by something called the boundary entropy. This is directly related to a renormalization gruop flow of the boundary CFT. 

In this context we will study the QRRE in an infinite fermionic chain with a defect on its boundary. We will show numerically how this quantities depends on strength of the defect through the effective central charge of the theory and how they depend on the size of the subsystem. The method to use will be just diagonalization of correlators and nothing more involved than that.

In section \ref{sec2} we will review the model under consideration and we will perform the computation of the entanglement entropy for a subsystem with the defect on its boundary. In section \ref{sectQRRE} we will define the quantum relative Renyi entropies and its connection with the quantum fidelity and the relative entropy. Moreover we will perform the numerical analysis. Lastly in section \ref{conclu} we will summarize the results and open a window to some related problems that will be addressed in forthcoming works.

\section{Infinite chain with and interface defect}\label{sec2}

In this section we will set the model under study and review the computation of the entanglement entropy done in \cite{ingo} because this is going to be related with our results for the QRRE but many work was done on the topic of entanglement entropy and defects, for example in \cite{sorensen, aff, sale}.

Consider a free spinless fermions hopping between neighboring sites of an infinite linear chain. The Hamiltonian is written as

\begin{equation}
{\cal H}=-\sum_n t_n\left(c_nc_{n+1}^\dag+c_{n+1}^\dag c_n\right)+\sum_n\Delta_n c_n^\dag c_n,\label{Hamiltonian}
\end{equation}
with $t_n$ the hopping matrix element. We will use $t_n=1$ for $n\neq 0$ and $t_0=t$. Moreover at first step we will use $\Delta_n=0$ and just in section \ref{delta} we will use $\Delta_1=\Delta$ and $t_n=1$ for all $n$ . In the present analysis we will put the defect on the boundary of the subsystem $A$. The number of sites on $A$ will be the length of the subsystem and we will denote it by $L$. The reduced density matrix is obtained by tracing over the degrees of freedom of the region complementary to $A$. In the present case the states are Gaussian and we can write them as
\begin{equation}
\rho_A= e^{-\sum_{i,j}H_{i,j}^A c_i^\dag c_j},\label{density}
\end{equation}
where $H$ is the entanglement Hamiltonian, which can be written as function of the correlators \cite{peschelold}

\begin{equation}
H_A=\log[C_A^{-1}-1],\label{EH}
\end{equation}
with $C$ the matrix
\begin{equation}
C_{i,j}^A=\langle c_i^\dag c_j\rangle, \,\,\,\,\,\, i,j\subset A.
\label{correl}
\end{equation}
Along the whole work we will be treating with reduced density matrices and quantities that will be computed from them, then we will drop the subscript $A$ in the rest of the manuscript.
For a homogeneous infinite system the correlators have a closed translational invariant form 
\begin{equation}
C_{ij}=C^0(i-j)=\frac{\sin(\pi(i-j)/2)}{\pi(i-j)}\label{C0}.
\end{equation}
For single defects the translational invariance is lost and we have
\begin{equation}
C_{ij}=C^0(i-j)-C^1(i+j)\label{C}.
\end{equation}
When we have a single weak bond $t=e^\nu\leq 1$ we obtain the following form for the correlators $C^1(l)$ \cite{ingo}
\begin{equation}
C^1(l)=-\frac12\sinh(\nu)\left(e^{-\nu}I_l-e^\nu I_{l-2}\right),\,\,\,\,\,\,\, I_l=\int_0^{\frac{\pi}{2}}\frac{dq}{2}\frac{\cos(q l)}{\sinh^2\nu +\sin^2q}\label{C1}
\end{equation}
For $t=0$, $C^1$ has the same form as $C^0$ but using $i+j$ instead of $i-j$. Then, in this case the correlators are those for a system with an open end. By diagonalizing  \eqref{C} we obtain the eigenvalues $\zeta_k$ ($0<\zeta_k<1$) from which the eigenvalues of H, $\epsilon_k$, can be read
\begin{equation}
\epsilon_k=\log(\zeta_k^{-1}-1).
\end{equation}
The study of the entanglement Hamiltonian  \eqref{EH} through its eigenvalues as function of the defect is an interestig issue by itself (see \cite{Arias, peschelold, chm, ct, ariaseh} for it study in different setups). The entanglement entropy in terms of this spectrum can be written as

\begin{equation}
S=- Tr\left(\rho\log\rho\right)=\sum_k \left(\zeta_k \log\zeta_k+(1-\zeta_k)\log(1-\zeta_k)\right).\label{EE}
\end{equation}

In \cite{ingo} the authors showed that the entanglement entropy as function of the defect strength and the size of the subsystems can be written in the following way
\begin{equation}
S=\frac{c_{eff}}{3}\log L+k,
\label{SvsLt}
\end{equation}
 where the effective central charge is \cite{ingo2, ingo3, pasquale2}
 \begin{equation}
 c_{eff}(t)=\frac{1}{2}+\frac{1}{2}{\cal C}(t),
 \label{ceff}
 \end{equation}
with
\bea
 {\cal C}(t)&=&-\frac{6}{\pi^2}\left[\left((1+s)\log(1+s)+(1-s)\log(1-s)\right)\log s\right.\nonumber\\&&\left.+(1-s) Li_2(-s)+(1-s) Li_2(s)\right], 
 \eea
and $s=\frac{2}{t+1/t}$. The parameter $s$ is the transmission coefficient through the defect at the Fermi level. The function $Li_2$ is the dilogarithm function defined as
\begin{equation}
Li_2(z)=-\int_0^z dx\,\, \frac{\log(1-x)}{x}\nonumber.
\end{equation}
Note that the $c_{eff}$ coefficient goes smoothly from $c_{eff}=1/2$ when $t=0$ to $c_{eff}=1$ when $t=1$.

\begin{figure}[t!] \label{ fig7} 
\hspace{-1.6cm}
  \begin{minipage}[c]{0.6\linewidth}
  \begin{center}  
    \includegraphics[width=0.95\linewidth]{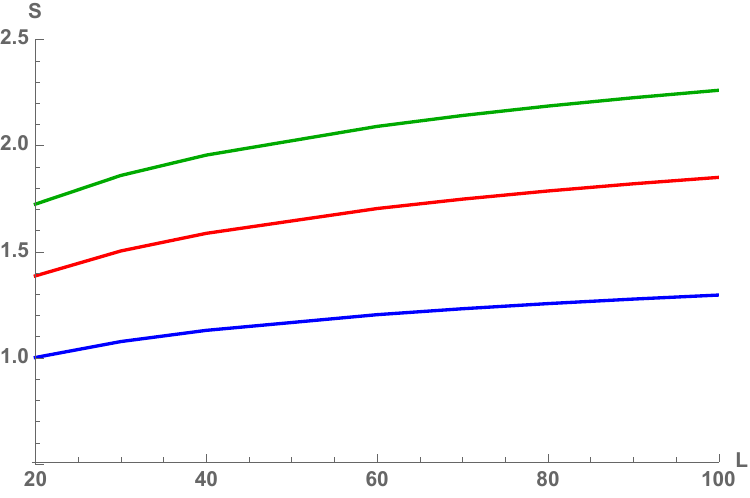} 
    \caption{Entanglement Entropy of the subsystem $A$ as function of its size $L$. The blue curve is for $t=0.1$, red curve for $t=0.5$ and green for the homogeneous $t=1$ case.} 
   \label{fig:SvsL} 
   \end{center}
  \end{minipage} 
  \begin{minipage}[c]{0.62\linewidth}
    \includegraphics[width=0.95\linewidth]{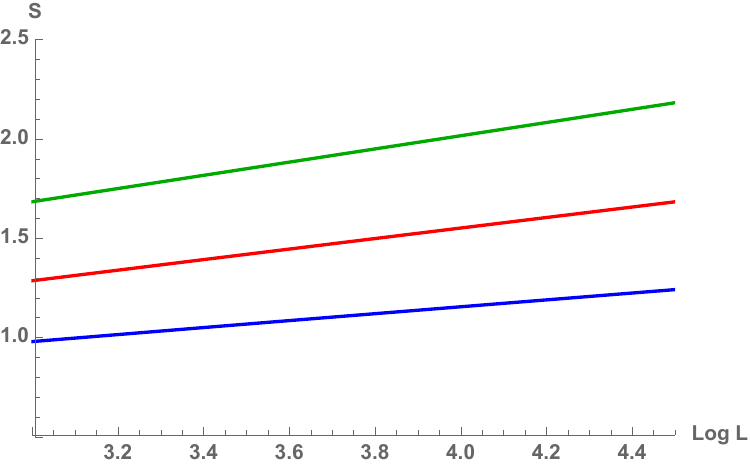} 
    \caption{Entanglement Entropy of the subsystem $A$ as function of $\log L$ for the same values of the defect than in the figure on the left. } 
     \label{fig:SvslogL} 
  \end{minipage} 
\end{figure}

\begin{figure} 
 \hspace{-1.6cm}
\begin{minipage}[c]{0.6\linewidth}
  \begin{center}
    \includegraphics[width=0.95\linewidth]{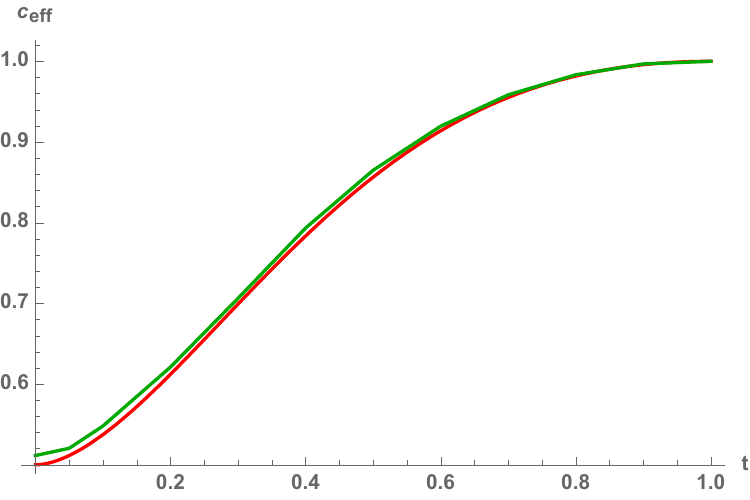} 
    \caption{Effective central charge as function of $t$. The red line is the analytic function \eqref{ceff} and the green line the numerical result.} 
     \label{fig:ceffvst} 
    \end{center}
  \end{minipage}
  \begin{minipage}[c]{0.6\linewidth}
    \includegraphics[width=0.95\linewidth]{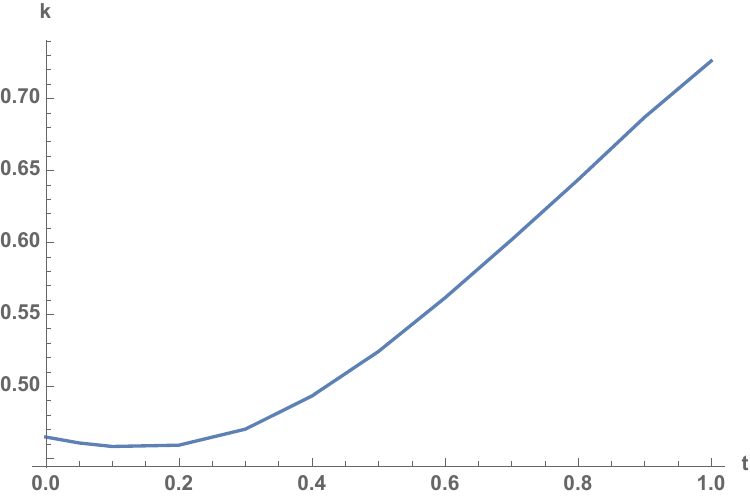} 
    \caption{Behavior of the coefficient $k$ in \eqref{SvsLt} as function of the defect strength $t$. } 
     \label{fig:kvst} 
  \end{minipage} 
\end{figure}

In figures \ref{fig:SvsL} and \ref{fig:SvslogL} we showed the behaviour of the entanglement entropy as function of $L$, and $\log L$ respectively, for three different values of the defect $t$, $t=0.1$ (in blue), $t=0.5$ (in red) and $t=1$ (in green). Moreover on figures \ref{fig:ceffvst} and \ref{fig:kvst} we showed the behaviour of the effective central charge and the coefficient $k$ in \eqref{SvsLt} as function of t. In particular in figure \ref{fig:ceffvst} we showed the analytic function \eqref{ceff} in red and the numerical result in green. 
Note the good agreement between our graphics and figures 4, 5 and 7 of \cite{ingo}.

\section{Quantum Renyi Relative Entropies}\label{sectQRRE}

In this section we will define the notion of quantum relative Renyi entropies and review some of its properties. After that we will show the numerical results when we compute these quantities in the fermion chain model of the previous section. 

\subsection{Definition and properties}

The quantum Renyi relative entropies between the state $\rho$ and the state $\sigma$ are defined as \cite{muller, wilde}
\begin{equation}
S_{\alpha}(\rho || \sigma)=-\frac{1}{1-	\alpha}\log\left[{\mathrm {Tr}}\left(\sigma^{\frac{1-	\alpha}{2\alpha}}\,\rho\, \sigma^{\frac{1-\alpha}{2\alpha}}\right)^\alpha\right], \label{QRREdef}
\end{equation}
for any value $\alpha \subset [1/2,1) \cup (1,\infty)$. When $\alpha=1/2$ the equations gives the quantum fidelity between the states 
\begin{equation}
S_{1/2}(\rho||\sigma)=-2\log F(\rho,\sigma),\label{fidelity}
\end{equation}
where we called $F$ to the fidelity function. In physical terms this is another measure of distance between the states $\rho$ and $\sigma$. In terms of wavefunctions the quantum fidelity is just the modulus of the overlap (Uhlmann theorem \cite{uhlmann})
\begin{equation}
F(\psi, \psi')=|\langle \psi|\psi'\rangle|,\label{uhlmann}
\end{equation}
where the wavefunction $|\psi\rangle$ is the purification associated to the density matrix $\rho$ and $|\psi'\rangle$ is the one associated to $\sigma$. The fidelity is symmetric in  its inputs $\rho,\sigma$ and is bounded $0\leq F(\rho,\sigma)\leq 1$ being $F(\rho,\sigma)=1$ when $\rho=\sigma$ and zero if and only if the states lives on orthogonal spaces. From this property is maybe more clear why it can be used to find critical points along quantum phase transitions (see \cite{gu}).

On other hand the limit $\alpha \rightarrow 1$ gives the Relative entropy, which is a measure of distinguishability between the states,

\begin{equation}
S_{\alpha \rightarrow 1}(\rho||\sigma)=Tr(\rho\log\rho-\rho\log\sigma)=S(\rho||\sigma).
\end{equation}
This quantity has the benefit that can be written in terms of another two quantum information quantities defined in the previous section, the entanglement entropy and the entanglement Hamiltonian associated to the reference state $\sigma$ ($H_\sigma$)
\begin{equation}
S(\rho||\sigma)=-\Delta S+\Delta\langle H_\sigma\rangle,\label{relaqft}
\end{equation}
with $\Delta S=S(\rho)-S(\sigma)$ and $\Delta\langle H_\sigma\rangle= \langle H_\sigma\rangle_{\rho}-\langle H_\sigma\rangle_{\sigma}$. This form of the relative entropy is mostly used in quantum field theory.  In general, the entanglement Hamiltonian is a complicated object because in many situations is a highly non local object. Although, knowing $\Delta S$ and the relative entropy one can obtain the result for the difference of it expectation value between the states.

Then, equation \eqref{QRREdef} is a family of distance measures that interpolates between the quantum fidelity and the relative entropy when $1/2\leq\alpha<1$ and because of this it is going to be the range of $\alpha$ in which we are interested in along this work. As function of $\alpha$ the equation \eqref{QRREdef} has some properties, it is a monotonically increasing function, it is greater or equal than 0 and is monotonically increasing when we increase the algebra of operators, or in other words when we increase the size of the region. In equations these three properties read:
\bea
\frac{d S_{\alpha}(\rho||\sigma)}{d\alpha}\geq 0,\nonumber\\
S_{\alpha}(\rho||\sigma)\geq 0,\,\,\, S_{\alpha}(\rho||\sigma)=0 \,\,\,\,\,{\mathrm{iff}}\,\,\,\,\, \rho=\sigma,\nonumber\\
S_{\alpha}(\rho_V||\sigma_V)\leq S_{\alpha}(\rho_{V'}||\sigma_{V'}),\,\,\ V \subseteq V'.\label{proper}
\eea

Then, our strategy is compute the quantum Renyi relative entropies between a homogeneous state ( i.e. $t=1$) and a different state with a value of the defect on the range $0\leq t\leq 1$.  We are going to take advantage on the fact that it was shown that the expression \eqref{QRREdef} can be written in terms of the correlation matrices of the states (see appendix A of \cite{horacio} for a carefully derivation). The equation reads
\begin{eqnarray}\label{QRRE}
S_\alpha(\rho || \sigma)&=&-Tr\log(1-C)-\frac{\alpha}{1-\alpha}Tr\log(1-C')\\&&-\frac{1}{1-\alpha}Tr\log\left[1+\left(\left(\frac{C}{1-C}\right)^{\frac{1-\alpha}{2\alpha}}\left(\frac{C'}{1-C'}\right)\left(\frac{C}{1-C}\right)^{\frac{1-\alpha}{2\alpha}}\right)^\alpha\right]\nonumber,
\end{eqnarray}
where $C$ is the correlation matrix of the state $\sigma$ and $C'$ is the correlation matrix of state $\rho$.

\subsection{Numerical results}

In this section we will show that the numerical analysis lead to a closed equation for the quantum Renyi relative entropies. The strategy is diagonalize the correlators for both states and use equation \eqref{QRRE} to obtain the results. Then, with the behaviour of $S_\alpha$ as function of $t$ at hand we will do a guess for $S_\alpha(t)$ as function of $L$ and for many values of $\alpha$.
 
 \subsubsection{Defect on the boundary}\label{defect}
  
\begin{figure}[t!]
\hspace{-1.6cm} 
 \begin{minipage}[c]{0.6\linewidth}
  \begin{center}  
\includegraphics[width=.95\textwidth]{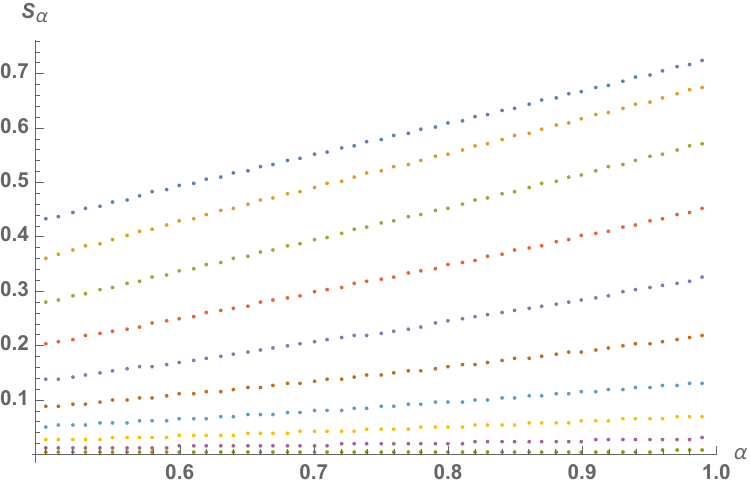}
 \end{center}
\caption{$S_{\alpha}(\rho ||\sigma)$ as function of $\alpha$ for $L=100$ and many values of the defect $t$ between 0 and 0.9 from top to bottom.}
\label{fig:SalphavsalphaL100}
\end{minipage}
\begin{minipage}[c]{0.61\linewidth}
  \begin{center}
  \includegraphics[width=0.95\textwidth]{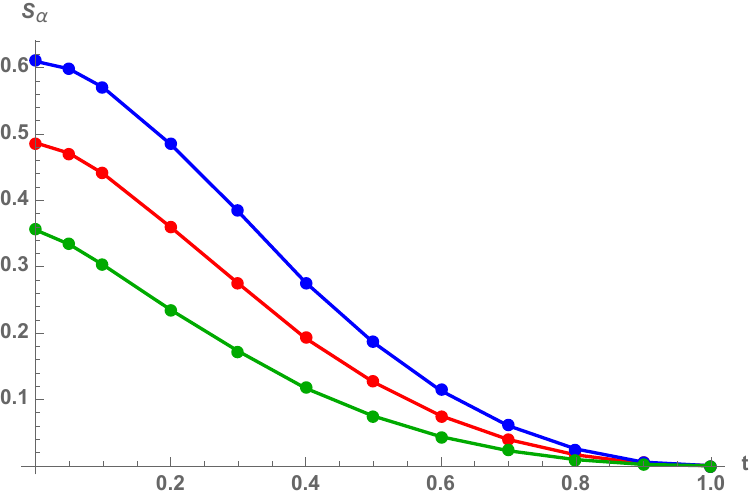}
\end{center}
\caption{$S_{\alpha}(\rho ||\sigma)$ fot three different values of $\alpha$ and $L=100$. In blue the relative entropy, $S_{0.99}$, in red $S_{0.75}$ and in green the one related to the fidelity, $S_{0.5}$ }
\label{fig:trealphavstL100}
\end{minipage}
\end{figure}

Figure \ref{fig:SalphavsalphaL100} shows how is the dependence of $S_\alpha(\rho||\sigma)$ in $\alpha$ for different values of the defect $t$, the figure is done for $L=100$. Note that the curves are monotonically increasing with $\alpha$  (see first equation of \eqref{proper}) and that the slope is decreasing and goes to 0 when $t$ goes to 1, which is expected because of the second property in \eqref{proper}. 

Figure \ref{fig:trealphavstL100} shows an example of how $S_{1/2}$ (in green), $S_{3/4}$ (red) and $S_{0.99}$ (blue) behaves as function of  the defect $t$. As is expected they vanish for $t=1$ because both inputs in $S_\alpha$ in that case are the same. 

\begin{figure}[t!]
\hspace{-1.6cm} 
 \begin{minipage}[c]{0.6\linewidth}
  \begin{center}  
\includegraphics[width=.95\textwidth]{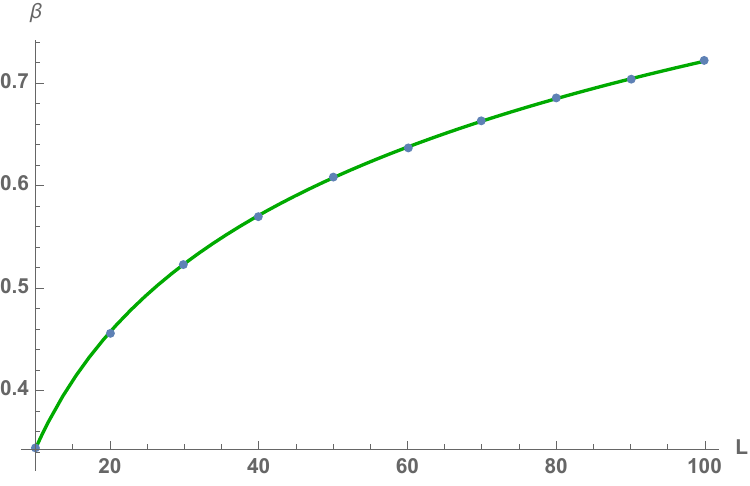}
 \end{center}
\caption{Coefficient $\beta$ in equation \eqref{relacoeffi} as function of $L$. In green the analytic function and the points are the numerical data.}
\label{fig:betavsL}
\end{minipage}
\begin{minipage}[c]{0.6\linewidth}
  \begin{center}
  \includegraphics[width=0.95\textwidth]{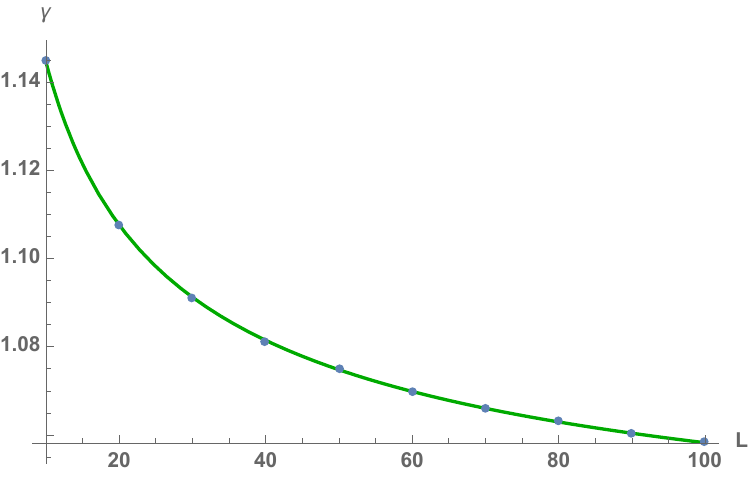}
\end{center}
\caption{Coefficient $\gamma$ in equation \eqref{relacoeffi} as function of $L$. In green the analytic function and the points are the numerical data.}
\label{fig:gammavsL}
\end{minipage}
\end{figure}

The behavior of $S_{0.99}$ remember us the behavior of the function ${\cal C}(t)$ involved in the effective central charge equation \eqref{ceff}. Then, one can try to fit a function on top of the numerical data which vanishes at $t=1$. The proposed function for the $\alpha=0.99$ case is
\begin{equation}
S_{0.99}=\beta\left(1- \left(2c_{eff}(t)-1\right)^\gamma\right).
\label{relaconje}
\end{equation}
For fixed $L$ and $\alpha$ one can obtain the coefficients $\beta$ and $\gamma$ in order to obtain a perfect agreement with the numerical data. Once the dependence on the defect is fixed one can see how is the behavior of the coefficients as function of the length. Again for the $\alpha=0.99$ case this goes as
\begin{equation}
\beta(L)=\frac{b_{0.99}}{3}\log L+d_{0.99},\,\,\,\,\,\,\,
\gamma(L)=a_{0.99}+ c_{0.99} L^{-1/2}.\nonumber\\
\label{relacoeffi}
\end{equation}
Figures \ref{fig:betavsL} and \ref{fig:gammavsL} show the numerical data for $\beta$ and $\gamma$ and the fitted functions, showing a perfect agreement.

\begin{figure}[t!] \label{ fig7} 
\hspace{-1.6cm} 
  \begin{minipage}[c]{0.6\linewidth}
  \begin{center}  
    \includegraphics[width=.95\linewidth]{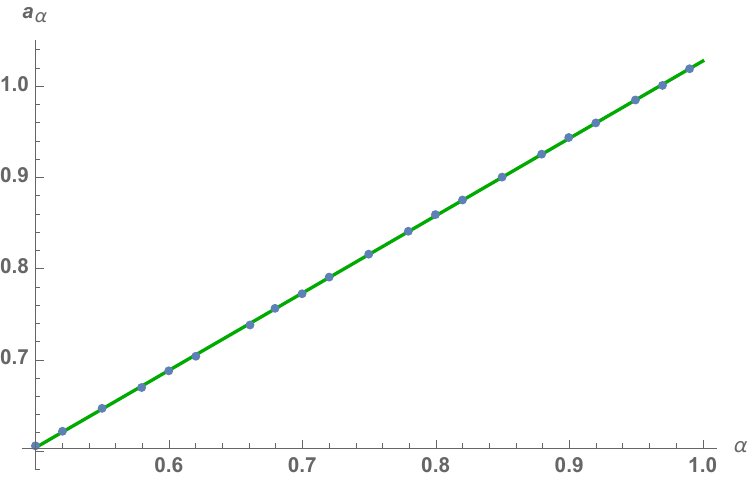} 
    \caption{Coefficient $a_\alpha$ in equation \eqref{relacoeffi} as function of $\alpha$. In green the fitted function in \eqref{coeffs}} 
   \label{fig:avsalfa} 
   \end{center}
  \end{minipage} 
  \begin{minipage}[c]{0.62\linewidth}
    \includegraphics[width=.95\linewidth]{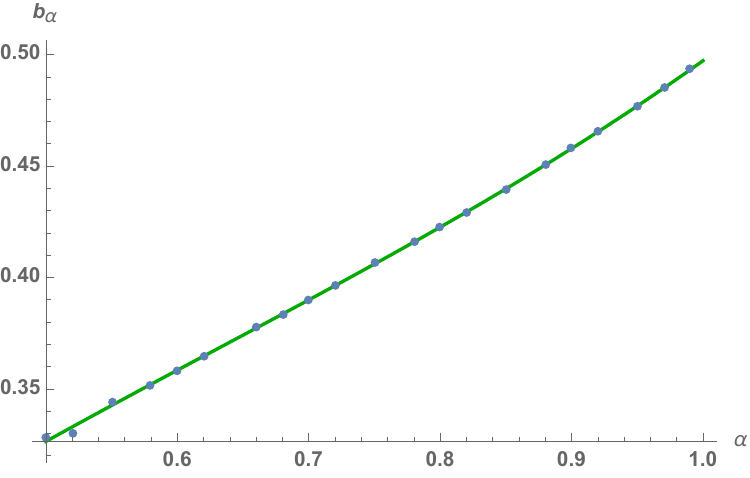} 
    \caption{Coefficient $b_\alpha$ in equation \eqref{relacoeffi} as function of $\alpha$. In green the fitted function in \eqref{coeffs}. } 
     \label{fig:bvsalfa} 
  \end{minipage} 
  \end{figure}

\begin{figure}
  \hspace{-1.6cm} 
  \begin{minipage}[c]{0.6\linewidth}
  \begin{center}
    \includegraphics[width=.95\linewidth]{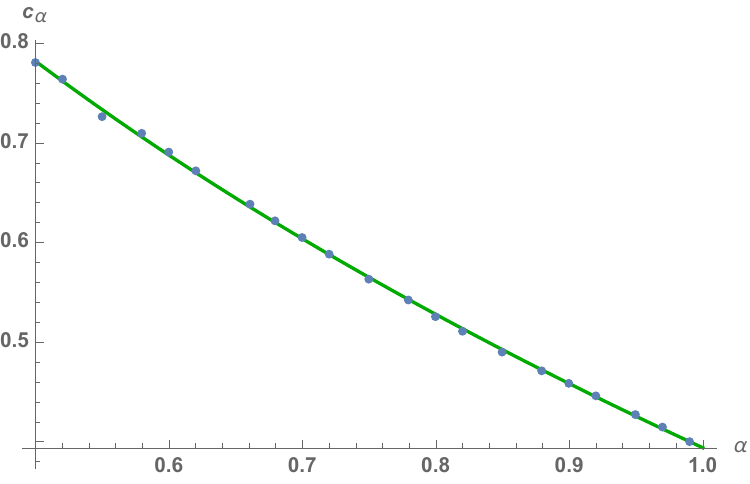} 
    \caption{Coefficient $c_\alpha$ in equation \eqref{relacoeffi} as function of $\alpha$. In green the fitted function in \eqref{coeffs}.} 
     \label{fig:cvsalfa} 
    \end{center}
  \end{minipage}
  \begin{minipage}[c]{0.6\linewidth}
    \includegraphics[width=.95\linewidth]{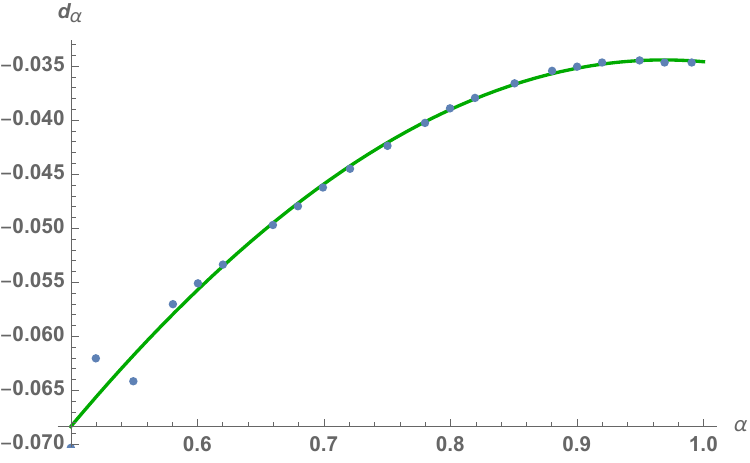} 
    \caption{Coefficient $d_\alpha$ in equation \eqref{relacoeffi} as function of $\alpha$. In green the fitted function in \eqref{coeffs}. } 
     \label{fig:dvsalfa} 
  \end{minipage} 
\end{figure}

One can see by numerical inspection that for any value of $\alpha$ an equation like \eqref{relaconje} can be written and then by studying the coefficients $a_\alpha,b_\alpha,c_\alpha,d_\alpha$ as function of $\alpha$ we can arrive to the general formula for the quantum Renyi relative entropies for this problem

\begin{equation}
S_{\alpha}\left(\rho||\sigma\right)= \left( \frac{b_\alpha}{3} \log(L)+d_\alpha\right) \left(1-\left(2c_{eff}(t)-1\right)^{a_\alpha+c_\alpha L^{-1/2}}\right),
\label{result}
\end{equation}
where $c_{eff}(t)$ was defined in \eqref{ceff} and the coefficients $b_\alpha, d_\alpha, a_\alpha, c_\alpha $ are polynomial functions of $\alpha$:
\bea
a_\alpha&=&0.847619\, \alpha +0.17972,\nonumber\\
b_\alpha&=&0.255266 \,\alpha ^3-0.481121 \,\alpha ^2+0.616596 \,\alpha +0.106444\nonumber\\
c_\alpha&=&2.46172\, -2.06784 \,\alpha ^{0.3},\nonumber\\
d_\alpha&=&-0.0232422 \,\alpha ^3-0.0978656 \,\alpha ^2+0.254972 \,\alpha -0.168467.
\label{coeffs}
\eea
Figures \ref{fig:avsalfa} - \ref{fig:dvsalfa} shows the agreement between the numerical data and the functions in \eqref{coeffs} (green full line).

Then, equation \eqref{result} tell us that all the QRRE goes as the logarithm of the subsystem size for large $L$. And this fact can be observed in equation \eqref{relaqft} for the relative entropy. In that expression the relative entropy is written as the difference of entanglement entropies with the form \eqref{SvsLt} that, in fact, are logarithmic. Then, it is not surprising that a logarithmic behaviour appeared in \eqref{result}. With this result at hand we can say that for this system $\Delta \langle H_\sigma\rangle \sim \phi(t) + \varphi(t) \log(L)$ where the coefficients $\phi$ and $\varphi$ depends on $t$ but they are in the range $-0.30008\leq\phi(t)\leq0,\,\,-0.0009 \leq\varphi(t)\leq0$ when $0\leq t \leq 1$.

Also note that in the limit $L\rightarrow \infty$ the fidelity \eqref{fidelity} goes to zero, and this can be seen as an expression of Anderson's orthogonality catastrophe \cite{anderson}. In this case the state of the system with the impurity is becoming pure and then the fidelity is given by \eqref{uhlmann}.


\begin{figure}[t] 
  \hspace{-1.6cm} 
  \begin{minipage}[c]{0.6\linewidth}
  \begin{center}  
    \includegraphics[width=.95\linewidth]{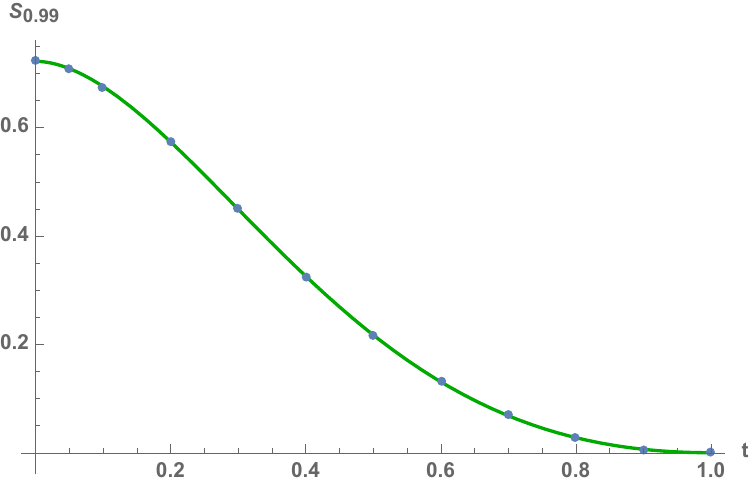} 
    \end{center}
    \caption{Numerical data and analytic result for $S_{0.99}$ as function of the defect $t$. The continues curve is equation \eqref{result} for $L=100$.} 
   \label{fig:ConjectureL100} 
\end{minipage}  
  \hfill
\begin{minipage}[c]{0.6\linewidth}
 \begin{center}  
    \includegraphics[width=.95\linewidth]{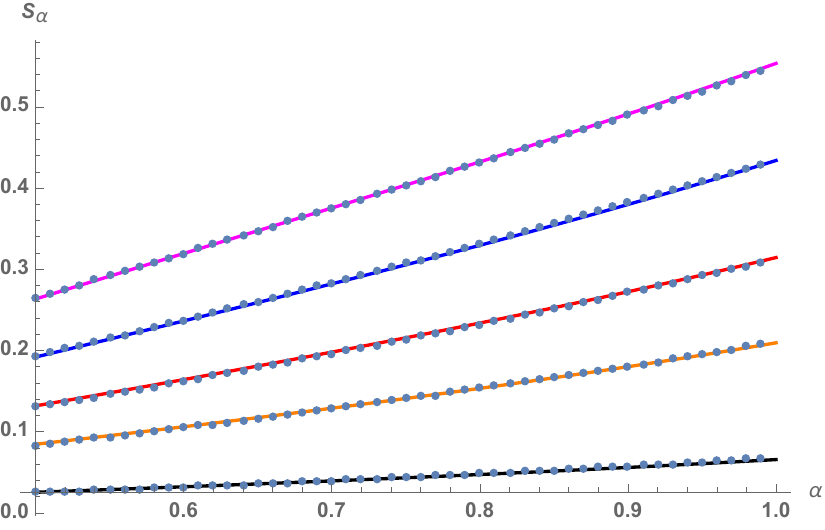} 
\caption{The figure shows $S_{\alpha}$ taken for various values of $t$ and $L=100$. The continuous curves are drawn using equation \eqref{result} and the points are the numerical data.} 
   \label{fig:Svsalfat}     
   \end{center}
   \end{minipage}
 \end{figure}

In figure \ref{fig:ConjectureL100} we show the perfect agreement for $S_{0.99}$ and $L=100$ between the numerical data and equation \eqref{result}. Lastly, in figure \ref{fig:Svsalfat} we see for fixed $L=80$ how the function \eqref{result} fits very well for different values of t as function of $\alpha$. Points are numerical data and full lines the analytic result for $t=0.2$ (in magenta), $t=0.3$ (in blue), $t=0.4$ (in red), $t=0.5$ (in orange) and $t=0.7$ (in black). 

\subsubsection{One site defect $\Delta\neq0$}\label{delta}

In this section we will set $t_n=1$ for any value of $n$ and $\Delta_1=\Delta$ next to the boundary. We will take the reference state for the homogeneous case $\Delta=0$ and we will compute the quantum Renyi relative entropies comparing it with another states for different values of $\Delta=2\sinh\nu$ between 0 and 10. 

In reference \cite{ingo} it was shown that the entanglement entropy has the same form as in equation \eqref{EE} but in this case there is no analytic form for the effective central charge. Then, on the numerical analysis we will proceed to obtain the QRRE by using an interpolating function for the effective central charge. The correlators for this theory that must be used to compute the physical quantities of interest are now
\begin{equation}
C^1(l)=\frac{1}{2}\sinh \nu\left(I_{l-1}-I_{l-3}+2\sinh\nu\, I_{l-2}\right).
\end{equation}

\begin{figure}[t!]
  \hspace{-1.6cm} 
 \begin{minipage}[c]{0.6\linewidth}
  \begin{center}  
\includegraphics[width=.95\textwidth]{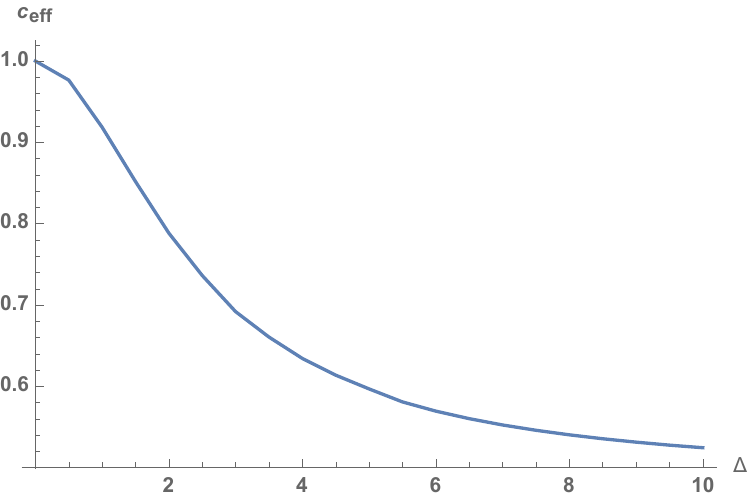}
 \end{center}
\caption{Effective central charge in equation \eqref{EE} as function of $\Delta$.}
\label{fig:ceffdelta}
\end{minipage}
\begin{minipage}[c]{0.6\linewidth}
  \begin{center}
  \includegraphics[width=0.95\textwidth]{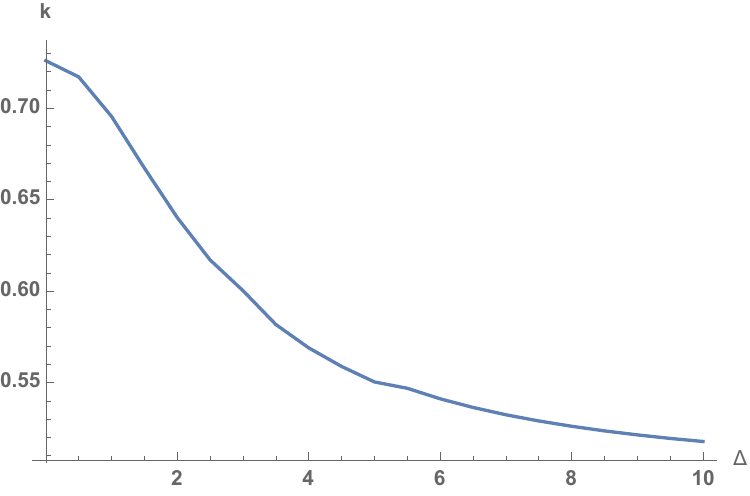}
\end{center}
\caption{Coeffcient $k$ in equation \eqref{EE} as function of $\Delta$.}
\label{fig:kdelta}
\end{minipage}
\end{figure}

 Doing the same procedure explained in section \ref{sec2} to compute the entanglement entropy we find the behaviour of $c_{eff}$ and $k$ as is shown in figures \ref{fig:ceffdelta} and \ref{fig:kdelta} respectively.  Note that, as in the previous case, the effective central charge interpolates between 1 for $\Delta=0$ and goes to 1/2 for large $\Delta$.

Performing the same analysis than for the previous case we found that 

\begin{equation}
S_{\alpha}\left(\rho||\sigma\right)= \left( \frac{b_\alpha}{3} \log(L)+d_\alpha\right) \left(1-\left(2c_{eff}(\Delta)-1\right)^{\gamma_\alpha(L)}\right),\label{resultdelta}
\end{equation}
where
\bea
b_\alpha&=&0.78053 \,\alpha^3-1.22826 \,\alpha^2+0.999614 \,\alpha-0.00375985,\\
d_\alpha&=&1.51467 \,\alpha^{0.8}-0.230572 .\label{coeffsdelta}
\eea

\begin{figure}[t!]
  \hspace{-1.6cm} 
 \begin{minipage}[c]{0.6\linewidth}
  \begin{center}  
\includegraphics[width=.95\textwidth]{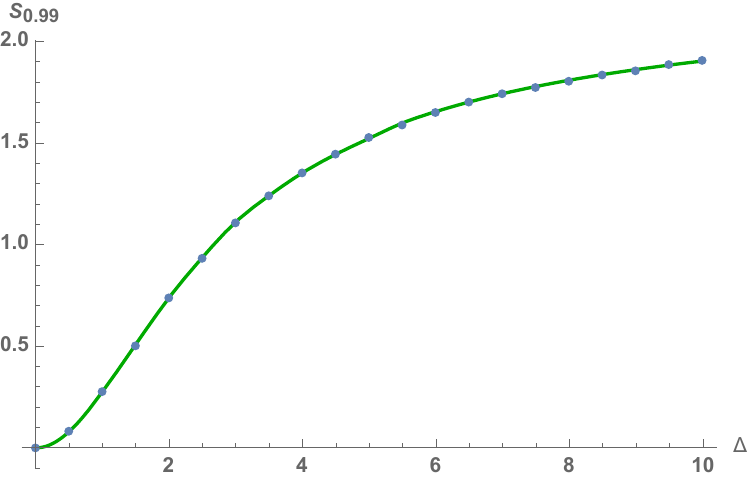}
 \end{center}
\caption{$S_{0.99}$ as function of $\Delta$. The continuous curve is the analytic result \eqref{resultdelta} with $\gamma_{0.99}$ taken from equation \eqref{gammaalfa}, the points are the numerical result.}
\label{fig:reldelta}
\end{minipage}
\begin{minipage}[c]{0.6\linewidth}
  \begin{center}
  \includegraphics[width=0.95\textwidth]{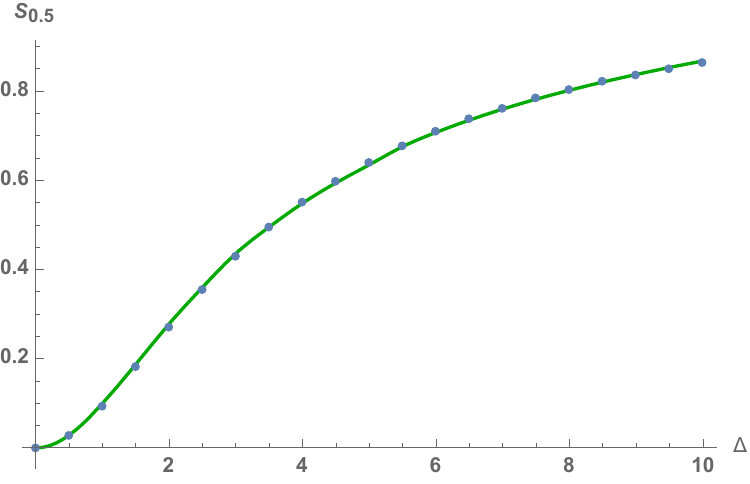}
\end{center}
\caption{$S_{0.5}$ as function of $\Delta$. The continuous curve is fthe analytic result \eqref{resultdelta} with $\gamma_{0.5}$ taken from equation \eqref{gammaalfa}, the points are the numerical result.}
\label{fig:fiddelta}
\end{minipage}
\end{figure}

Remarkably  we obtain the same function of the central charge as in the previous case but now  it is difficult to extract the $L$ dependence on the $\gamma$ coefficient because we do not have an analytic expression like \eqref{ceff} for the effective central charge. At this level we observe three different regimes as function of $L$, for $0.5<\alpha<0.62$  we can approximate $\gamma=c_0$ as a constant value. For $0.62<\alpha<0.85$ the fitted function is $\gamma=c_1 +c_2 \log(\log(L))$ and for $0.85<\alpha<1$ we have $\gamma=c_3 +c_4\log(L)$. Where the coefficients $c_i$ depends on the value of $\alpha$. The most interesting are the limiting cases which are related to the fidelity and the relative entropy, in that cases the coefficient $\gamma_\alpha(L)$ are
\bea
\gamma_{0.5} &=& 0.536104\nonumber\\
\gamma_{0.99} &=& 0.71725 +0.015165 \log(L).\label{gammaalfa}
\eea
Using these numbers we showed in figures \ref{fig:reldelta} and \ref{fig:fiddelta} how the numerical data and the equation \eqref{resultdelta} agrees perfectly. The figures shows the $\Delta$ dependence for $L=100$ but the agreement is as good as this for the other lengths we computed.

\section{Conclusions}\label{conclu}

In the present work we study the behaviour of the quantum Renyi relative entropies in a spinless fermionic model with interface defects in the lattice. First we review the model and the definition and properties of the QRRE, after that we compute $S_\alpha$ between a reference state corresponding to the homogeneous case and a second state corresponding to a non homogenous system. The main result is summarized in the equation \eqref{result} where one can see the dependence on the effective central charge of the theory. At this point is good to mention that in \cite{igloi} was shown that the $XX$ model eigenvalues $\epsilon_k$ are related to the ones in the transverse Ising chain and as consequence both systems have the same effective central charge. Then, one can guess that our result can be also applied to that case but is a good point to explore in the future.

Moreover we did the same procedure to study the site defect with energy $\Delta$ and found that the dependence on the effective central charge is the same.
One can then ask if this quantities depends on a central charge in a more general setup or if this is strongly model dependent. 

It is interesting to ask how this system with a defect is related to the boundary conformal field theory studied in \cite{horacio} and how the quantum Renyi relative entropies computed in our work are related with that ones, if there is a connection between the two models.

Moreover one can extend our analysis to the two defect case studied also in \cite{ingo}. We leave for future work a more interesting situation in a time dependent setup as those studied in \cite{LocalQ, igloi2} where the authors analyze the behavior of the entanglement entropy for free electrons on a half-filled infinite chain with a bond defect after a quench.

\section*{Acknowledgments}
The author want to thank to Raimel Medina Ramos, Ingo Peschel, Erik Tonni and Gonzalo Torroba for valuable discussions.

\end{document}